\documentclass[prb,aps,twocolumn,superscriptaddress]{revtex4-1}
\usepackage{graphicx,color}
\usepackage{amsthm}
\usepackage{amsfonts}
\usepackage{algorithmic}
\usepackage{enumerate}
\usepackage{latexsym}
\usepackage{amsmath}
\usepackage{amssymb}
\usepackage[colorlinks=true,citecolor=blue,linkcolor=blue]{hyperref}

\usepackage{ulem}

\emergencystretch=\maxdimen
\begin{document}

\title{Intermediate phase in interacting Dirac fermions with staggered potential}

\author{Jingyao Wang}
\affiliation{Department of Physics, Beijing Normal University, Beijing
100875, China}
\affiliation{Department of Physics and Astronomy, University of Southern California, Los Angeles, CA 90089-0484, USA}
\author{Lufeng Zhang}
\affiliation{School of Science, Beijing University of Posts and Telecommunications,
Beijing 100876, China}
\affiliation{Department of Physics, Beijing Normal University, Beijing
100875, China}
\author{Runyu Ma}
\affiliation{Department of Physics, Beijing Normal University, Beijing
100875, China}
\author{Qiaoni Chen}
\affiliation{Department of Physics, Beijing Normal University, Beijing
100875, China}
\author{Ying Liang}
\affiliation{Department of Physics, Beijing Normal University, Beijing
100875, China}
\author{Tianxing Ma}
\email{txma@bnu.edu.cn}
\affiliation{Department of Physics, Beijing Normal University, Beijing
100875, China}
\affiliation{Beijing Computational Science Research Center, Beijing
100193, China}

\begin{abstract}
By performing exact quantum Monte Carlo simulations of a model of interacting Dirac Fermions with staggered potential, we reveal a novel intermediate phase where the electronic correlations drive a band insulator metallic, and at a larger interaction, drive the metal to Mott insulator. We also show that the Mott insulating phase is antiferromagnetic.
A complete phase diagram is achieved by studying the phase transitions at large staggered potential and interaction strengths, which shows that the intermediate state is robust and occupies a large part of the phase diagram and that it should be more feasible to be detected experimentally.
\end{abstract}

\maketitle

\section{Introduction}
\label{sec:intro}
%\noindent
%\underline{\it Introduction}---
Since the discovery of graphene\cite{Novoselov2005,*Zhang2005,RevModPhys.81.109} and topological insulators\cite{RevModPhys.82.3045,*RevModPhys.83.1057},
Dirac fermions described by a honeycomb lattice, has enriched our knowledge of physics beyond Landau's symmetry breaking theory\cite{Peres2010,Mourik1003,Grover280}.
Landau's theory of the Fermi liquid describes the interacting electrons of a typical metal as an ideal gas of noninteracting quasiparticles.
This description is expected to fail for Dirac fermions due to their linearly dispersing bands and minimally screened Coulomb interactions\cite{RevModPhys.81.109}.
Half-filled graphene hosts a Dirac fluid governed by relativistic hydrodynamics\cite{Crossno1058, *Gallagher158}.
Inspired by recent experiments on twisted bilayer graphene\cite{Cao2018A,*Cao2018B,Yankowitz1059,HUANG2019310,*PhysRevB.101.155413},
a fast-growing frontier of research has focused on the novel physics induced by correlation effects in interacting Dirac fermions.

Correlation effects play an essential role in many intriguing physical properties of modern science, touching upon topics ranging from unconventional superconductivity\cite{RevModPhys.66.763,RevModPhys.78.17},
fractional quantum Hall effect\cite{Xu2009Fractional,Bolotin2009Observation},
quantum spin liquid\cite{Sorella2012,*PhysRevX.6.011029,RevModPhys.89.025003,PhysRevX.8.041040},
to metal-insulator phase transitions\cite{PhysRevLett.120.116601,PhysRevLett.98.046403}.
Those phenomena are all relevant to the ionic Hubbard model\cite{PhysRevLett.47.1750}, which contains the on-site Coulomb interactions and staggered potentials on bipartite lattices. Generally speaking, in a bipartite lattice like the honeycomb lattice, a broken inversion symmetry caused by an energy offset between the two sublattices, leads to a trivial band insulator at half filling, and the Coulomb interaction slows down the electrons or even localizes them in a Mott insulating phase, characterized by a spectral gap that opens\cite{Nature1476}.
Studies on the competition between the on-site Coulomb interactions and staggered potentials have witnessed extraordinary growth about possible exotic intermediate states between two or more competing phases.

This issue was very actively debated over more than a decade\cite{Egami1307,Nature1476,PhysRevLett.47.1750,PhysRevLett.74.4738,PhysRevLett.92.246405,PhysRevB.44.7143,PhysRevB.73.174516,PhysRevLett.98.016402,EBRAHIMKHAS20151053,PhysRevB.91.235108} before finally being settled\cite{PhysRevLett.115.115303,PhysRevLett.119.230403}.
The seminal work of dynamical mean-field theory (DMFT) studies on a square lattice\cite{PhysRevLett.97.046403} suggests that an interaction-induced metallic phase exists in the intermediate coupling region. Subsequently, cellular DMFT simulations have found a bond order phase\cite{PhysRevLett.98.016402} in this region, while determinant quantum Monte Carlo (DQMC) calculations of conductivity indicate a metallic phase\cite{PhysRevLett.98.046403,PhysRevB.76.085112}.
In addition to metallic and bond order insulating phases, various other phases depending on the lattice geometry have been proposed on other bipartite lattices, such as  charge-density-wave insulator\cite{PhysRevLett.98.016402}, superfluid\cite{PhysRevB.44.7143,PhysRevB.73.174516}, semimetal\cite{EBRAHIMKHAS20151053} and half-metal\cite{PhysRevB.91.235108}.
Recently, exciting progress on ultracold atom experiments has been made,
and the ionic Hubbard model was realized in an optical honeycomb lattice\cite{PhysRevLett.119.230403,PhysRevLett.115.115303}.
Unfortunately, only the band insulating phase and Mott insulating phase were observed.
Therefore, determining the existence of an intermediate phase or the nature of the intermediate phase is a subtle and largely open problem. Stimulated by the controversy and to motivate further experiments, in this paper,
we focus on the ionic Hubbard model on a honeycomb lattice, which is a minimal model that includes both interactions and staggered potentials in a two-dimensional (2D) Dirac system.
This model can not only be implemented in cold-atom systems but also can be realized on hydrogen graphene\cite{nmat2710}; moreover, the new family of 2D layered nitride materials Li$_x$\textit{M}NCl (\textit{M}=Hf, Zr) may be another candidate\cite{nature33362,*PhysRevLett.94.217002}.

\begin{figure}[t]
\centerline {\includegraphics[width=3in]{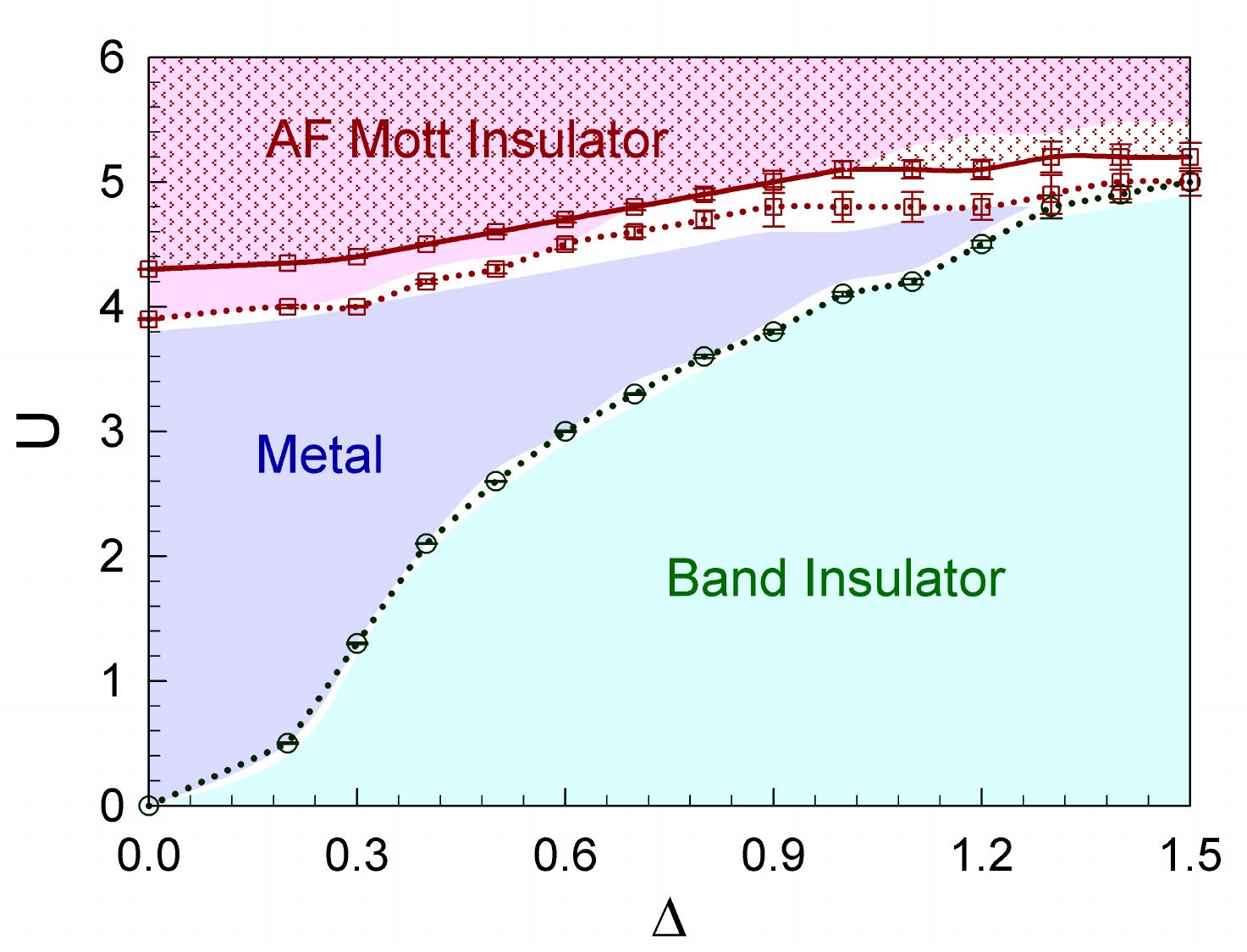}}
\caption{The phase diagram of the ionic Hubbard model on a $L$=12 honeycomb lattice. The phase boundaries are decided based on the temperature dependence of the conductivity $\sigma_{dc}$ and the finite-size scaling of the AFM structure factor.
Above the dark red solid line is where AFM order appears. The area covered with dark red dots above the dark red solid line is Mott insulating phase with AFM order.
The dark red dashed line represents transitions between Mott insulator and metal,
and the dark green dashed line indicates the phase boundary between metal and band insulator.
Those three colored areas distributed alongside the two transition
lines indicate different phases (green: band insulator, blue: metal,
pink: Mott insulator). The white color area indicates indistinct phases due to calculating errors.
}
\label{Fig:Phase}
\end{figure}

Our simulations were completed by the DQMC method on a half-filled case.
By varying the on-site interaction $U$, staggered potential $\Delta$, lattice size and temperature,
the bulk conductivity $\sigma_{dc}$ is calculated to determine either a metallic or an insulating phase,
and finite-size scaling is implemented to detect the long-range antiferromagnetic (AFM) order in the thermodynamic limit.
Our results reveal that an intermediate metallic state exists between the band and Mott insulators,
and the AFM order appears after the second metal-insulator transition with increasing $U$.
The exact numerical method that we are using successfully captures all the phase transitions at large-enough staggered potential and interaction strength,
and a complete phase diagram is achieved and shown in Fig. \ref{Fig:Phase}, which has several key differences relative to previous models.
First, the intermediate state that we found is more robust and occupies a larger part of the phase diagram.
For example, at small $\Delta$, the intermediate phase disappears quickly as $U$ increases in the ionic Hubbard model on a square lattice\cite{PhysRevLett.98.046403},
while here, the intermediate state is robust up to $U_c$=3.9. Second, the critical $U_c$ is in a reasonable range for experimental detection.
The intermediate insulator state in the square lattice vanishes around $U$=11\cite{PhysRevLett.98.016402} while it continues in the Haldane-Hubbard Model\cite{PhysRevLett.116.225305}. Furthermore, beyond previous results, we show a complete phase diagram where for large $\Delta$, the intermediate state disappears,
and the system transitions from band insulator to Mott insulator directly.

\section{Model and method}
\label{sec:model}
%\noindent
%\underline{\it Model and method}---

The Hamiltonian for interacting Dirac Fermions with staggered potential is
\begin{eqnarray}
\label{Hamiltonian}
\hat H=&-&t\sum_{{\bf i}\in A, {\bf j}\in B, \sigma}   (\hat c_{{\bf i}\sigma}^\dagger \hat c_{{\bf j} \sigma}^{\phantom{\dagger}}+H.c.)
+U \sum_{{\bf i}}  \hat n_{{\bf i}\uparrow} \hat n_{{\bf i}\downarrow} \nonumber \\
&+& \Delta\sum_{{\bf i}\in A,\sigma}\hat n_{{\bf i}\sigma}-\Delta\sum_{{\bf i}\in B,\sigma}\hat n_{{\bf i}\sigma} -\mu \sum_{{\bf i}\sigma} \hat n_{{\bf i}\sigma},
\end{eqnarray}
where $t$, $U$ and $\mu$ represent the nearest-neighbor electron hopping amplitude, on-site Coulomb repulsion, and chemical potential. The electron density of the system is characterized by the chemical potential $\mu$. $\hat c_{{\bf i}\sigma}^\dagger(\hat c_{{\bf i}\sigma}^{\phantom{\dagger}})$ is the operator that creates (annihilates) an electron with spin $\sigma$ at site ${\bf i}$, and $\hat n_{{\bf i}\sigma}$=$\hat c_{{\bf i}\sigma}^\dagger \hat c_{{\bf i}\sigma}^{\phantom{\dagger}}$. Specifically, $\Delta$ is the staggered one-body potential between sites in A and B sublattices with opposite signs. It is also called the ionic potential.
The band gap, $2\Delta$, has a nonzero value as a result of breaking the symmetry between sublattice A and B. We set $t$=1 as the default energy scale.

We adopt the exact DQMC method\cite{PhysRevD.24.2278,*PhysRevB.40.506} to study the phase transitions in the model defined by Eq.(\ref{Hamiltonian}).
DQMC is a powerful and reliable tool to investigate strongly correlated electron systems. In the DQMC method, the partition function $Z$=$\mathrm{Tr}\exp(-\beta H)$ is expressed as a path integral over the discretized inverse temperature $\beta$ over a set of random auxiliary fields. Then, the integration is accomplished by Monte Carlo techniques.
The on-site interaction is decoupled by a Hubbard-Stratonovich (HS) transformation, which leads to a sum over the discrete HS field and leaves the Hamiltonian in a quadratic form in the fermion operators. The resulting quadratic form can be integrated analytically and becomes the Boltzmann weight, expressed as the product of the determinants of two matrices that depend on the HS spin variables. In the half-filled ionic Hubbard model on the honeycomb lattice, the system is sign-problem free on account of the particle-hole symmetry, which allows us to achieve large $\beta$ simulations to converge to the ground state.

With the aim of exploring the phase transitions of the system, we computed the $T$-dependent DC conductivity, which is calculated from the wave vector $\textbf{q}$ and the imaginary time $\tau$-dependent current-current correlation function\cite{PhysRevB.54.R3756,PhysRevLett.75.312} $\Lambda_{xx}(\textbf{q},\tau)$:
\begin{eqnarray}
\label{DC}
\sigma_{dc}(T)=\frac{\beta^2}{\pi}\Lambda_{xx}(\textbf{q}=0,\tau=\frac{\beta}{2}) ,
\end{eqnarray}
where $\Lambda_{xx}(\textbf{q},\tau)$=$\left<\hat{j}_x(\textbf{q},\tau)\hat{j}_x(\textbf{-q},0)\right>$, $\beta$=$1/T$, $\hat{j}_x(\textbf{q},\tau)$ is the $(\textbf{q},\tau)$-dependent current operator in the $x$ direction. Eq.(\ref{DC}) has been employed for metal-insulator transitions in the Hubbard model in many works\cite{PhysRevLett.120.116601,PhysRevLett.75.312,PhysRevLett.83.4610}. A further way is to extract the spectral function by inverting the Laplace transform:
\begin{eqnarray}
\label{aomega}
G(\bf{q}=0,\tau)=\int d\omega \frac{e^{-\omega\tau}}{1+e^{-\beta\omega}A(\omega)},
\end{eqnarray}
in which $G(\bf{q}=0,\tau)$ can be achieved from the spatial Fourier transform of $G(\bf{R},\tau)=\langle c_{r+R\sigma}(\tau)c_{r\sigma}(0)\rangle$, and $A(\omega)$ is solved by performing with a method of analytic continuation.

We are also concerned about the magnetic properties of the system by studying the AFM spin structure factor\cite{Sorella2012,*PhysRevX.6.011029}
\begin{eqnarray}
\label{AFM}
S_{AFM}=\frac{1}{N_c}\left<\left(\sum\nolimits_{\textbf{r}\in A}\hat{S}^{z}_{\textbf{r}}-\sum\nolimits_{\textbf{r}\in B}\hat{S}^{z}_{\textbf{r}}\right)^2\right> ,
\end{eqnarray}
where $N_c$ represents the unit cell number of the lattice, $\hat{S}^{z}_{\textbf{r}}$ is the $z$ component of the spin structure factor operator, and the angle brackets $\left<\cdots\right>$ signify Monte Carlo simulations. To further study the nature of different stages of the system, we calculated the local moment $m$ by\cite{PhysRevB.87.125141}
\begin{eqnarray}
\label{eq:local moment}
m = \frac{1}{N_c}\sum_{\bf i}\left<(\hat{S}_{\bf i}^z)^2\right>
=1-\frac{2}{N_c}\sum_{\bf i}\left<\hat n_{{\bf i}\uparrow}\hat n_{{\bf i}\downarrow}\right>.
\end{eqnarray}

\begin{figure}[t]
\centerline {\includegraphics*[width=3.4in]{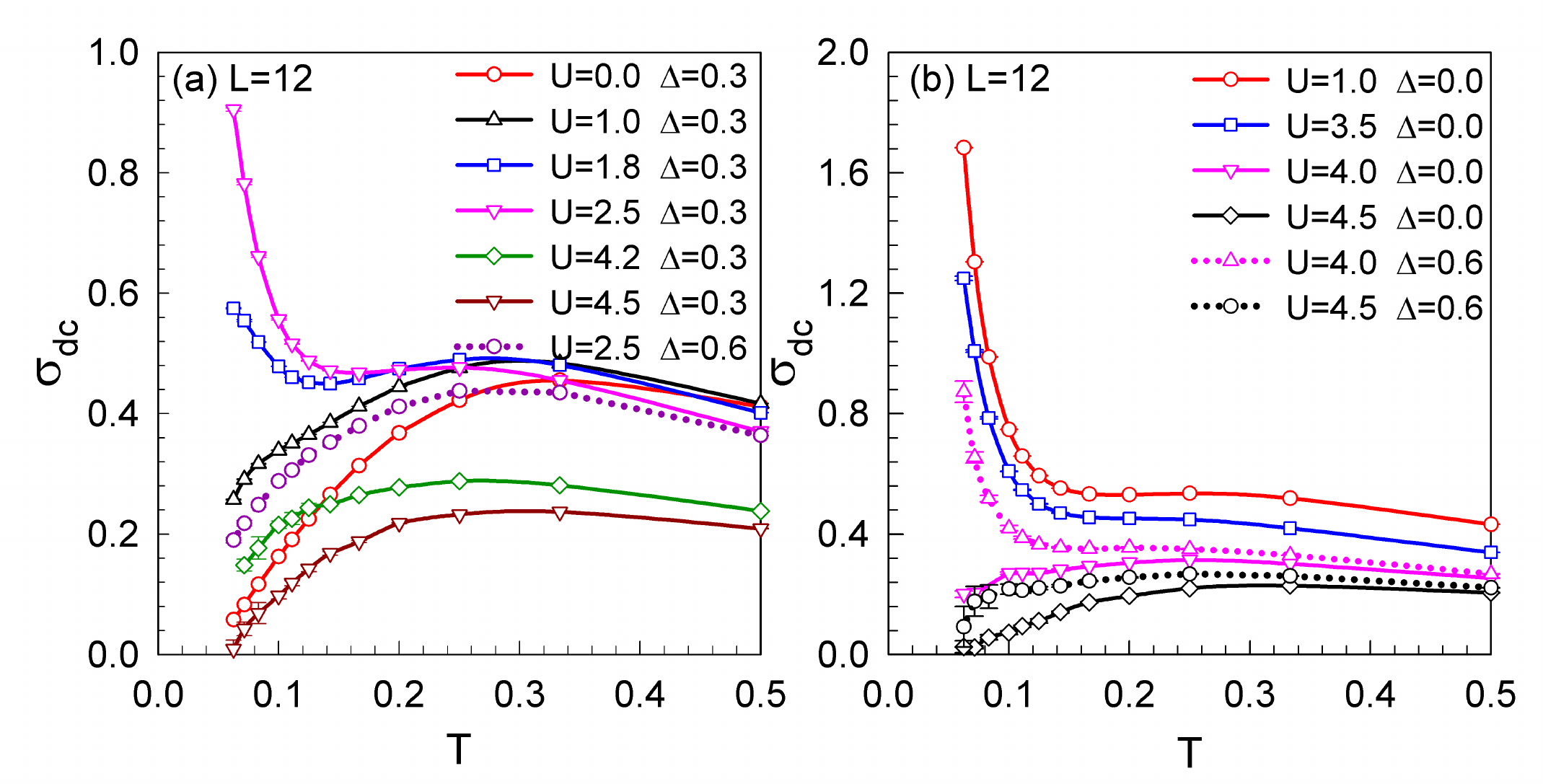}}
\caption{(a) (b) The conductivity $\sigma_{dc}$ is shown as a function of temperature at half-filling with increasing interaction $U$ for different $\Delta$. }
\label{Fig:UD}
\end{figure}

\section{Results and discussion}
\label{sec:results}

%\noindent
%\underline{\it Results and discussion}---
We first calculated the temperature dependence of conductivity $\sigma_{dc}$ with increasing interaction $U$ for a fixed value $\Delta$=0.3.
Basically, $d\sigma_{dc}/dT>0$ at low $T$ indicates an insulating phase, while $d\sigma_{dc}/dT<0$ at low $T$ corresponds to a metallic phase. In Fig.~\ref{Fig:UD} (a), we can see that at $U$=0.0 and $U$=1.0, the $\sigma_{dc}$ curve shows a concave down tendency and approaches the origin as the temperature decreases. This kind of low $T$ behavior suggests that the system exhibits insulating behavior. Interestingly, the behavior becomes metallic as the on-site interaction increases to $U$=1.8. A further increase in the interaction to $U$=2.5 reinforces the metallic behavior, but the advance to $U$=4.5 destroys the metallic state thoroughly and changes the system into a Mott insulator phase.
When the staggered potential increases to $\Delta$=0.6 for $U$=2.5, the increase in $\Delta$ suppresses the metallic behavior and insulating state develops, which also implies that the value of $U$=2.5 is not strong enough to drive the metallic phase for $\Delta$=0.6.

We present more data in Fig.~\ref{Fig:UD} (b) to emphasize the contrast and highlight the effect of the staggered potential.
For $\Delta$=0, at $U$=1.0 and $U$=3.5, $\sigma_{dc}$ increases as $T$ is lowered. When we calculate larger $U$ values ($U$=4.0 and $U$=4.5),
$\sigma_{dc}$ decreases as $T$ is lowered, which shows insulator behavior when the staggered potential is absent.
For the $U$=4.0 case, the insulating phase at $\Delta$=0 is displaced by a metallic phase when $\Delta$=0.6.

\begin{figure}[t]
\centerline {\includegraphics*[width=3.4in]{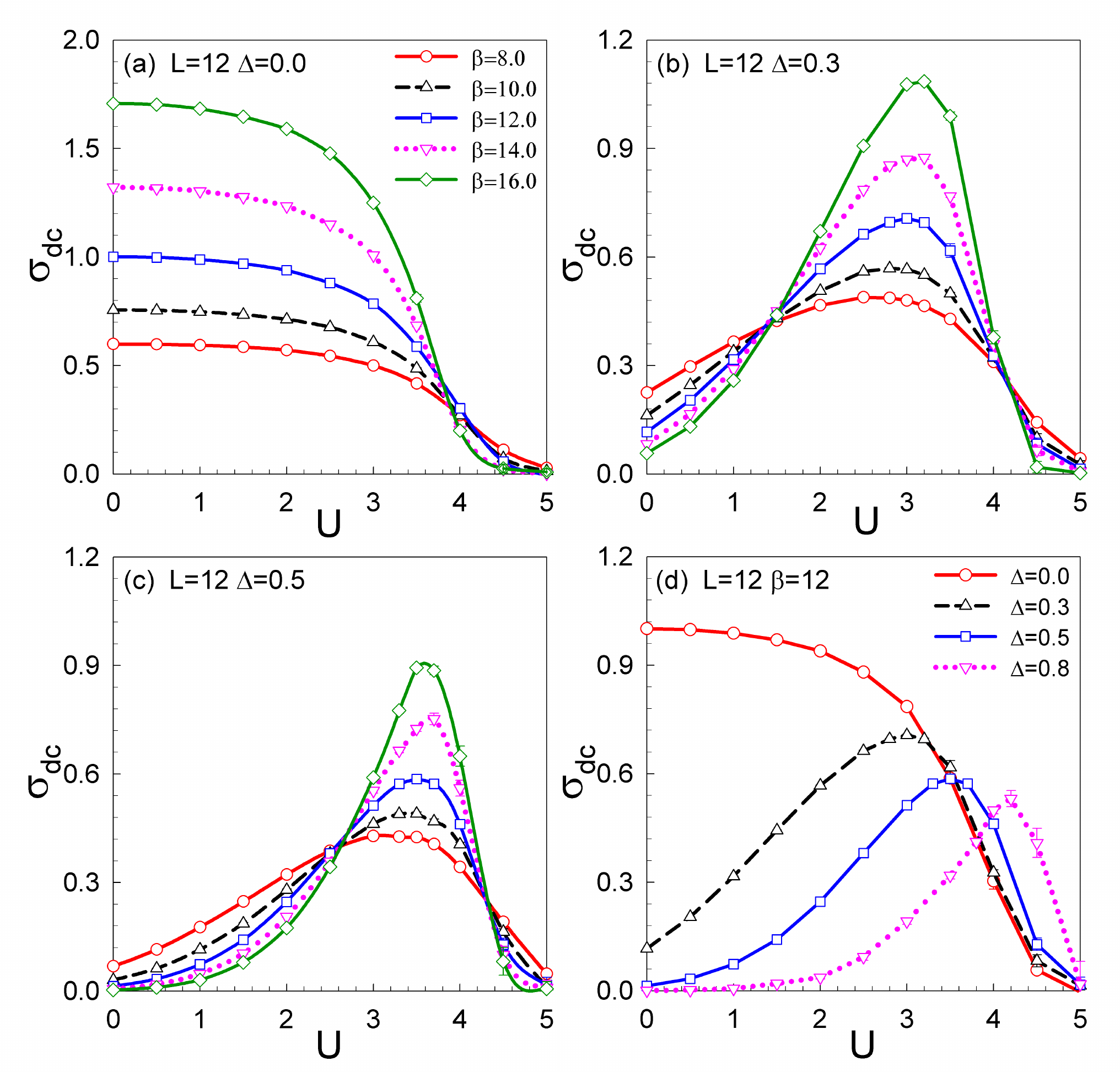}}
\caption{The conductivity $\sigma_{dc}$ at half-filling versus the interaction $U$ for different temperatures for (a) $\Delta$=0.0, (b) $\Delta$=0.3 and (c) $\Delta$=0.5. (d) The conductivity $\sigma_{dc}$ versus the interaction $U$ for different $\Delta$ at $\beta$=12 shows that the position of the largest conductivity moves to larger $U$ as $\Delta$ increases.}
\label{Fig:U}
\end{figure}

Fig.~\ref{Fig:UD} reveals the interesting phenomenon that the electronic correlation may drive a band insulator metallic, and at a larger interaction, there is a second transition from the metal to a Mott insulator.
To further explore this issue, we plotted Fig.~\ref{Fig:U}. Fig.~\ref{Fig:U}(a) shows that the transition from metal to Mott insulator is restored at $U_c$=3.9 for fixed $\Delta$=0.0.
In Fig.~\ref{Fig:U}(b), the curves intersect at two points, $U_{c1}$=1.4 and $U_{c2}$=4.2. In the range of 0$<$$U$$<$$U_{c1}$ and $U_{c2}$$<$$U$$<$5.0, the conductivity $\sigma_{dc}$ values at higher temperature exceeds those of lower temperature for the same $U$. The system maintains an insulating phase. The opposite situation emerges within the range of $U_{c1}$$<$$U$$<$$U_{c2}$. The conductivity increases as the temperature decreases, which demonstrates the metallic phase. The largest conductivity value for different $T$ values remains near $U_p$=3.0 at $\Delta$=0.3. The two crosspoints represent the transitions from band insulator to metal to Mott insulator.
Fig. \ref{Fig:U}(c) confirms these transitions with different $U_{c1}$=2.6 and $U_{c2}$=4.3 at $\Delta$=0.5, and the largest conductivity occurs at approximately $U_p$=3.5.

Interestingly, however, the position of the largest conductivity moves to larger $U$ as $\Delta$ increases, roughly following the law of $U_p(\Delta)$=2.5+2$\Delta$, as that shown in Fig. \ref{Fig:U}(d). This result is quite different from that of the ionic Hubbard model on a square lattice, where the largest conductivity remains near $U_p(\Delta)$=2$\Delta$, as one might expect from the $t$=0 analysis\cite{PhysRevLett.98.046403}.
In the Hubbard model on a square lattice, the charge density wave and local moments are perfectly balanced on the special $U$=2$\Delta$ line at $t$=0, and hence, the system is most likely to be metallic. At $t$=1, the AFM point also lies on that line, which is $U_c$=0 at $\Delta$=0, but for a honeycomb lattice, the AFM point lies much higher above
the line due to its Dirac fermion behavior at half filling. Therefore, perhaps the AFM point ``pulls'' the largest $\sigma_{dc}$ point up from $U$=2$\Delta$, and this ``pull'', also leaves a larger region of intermediate phase for interacting Dirac fermions with staggered potential.

\begin{figure}[htbp]
\centerline {\includegraphics[width=3in]{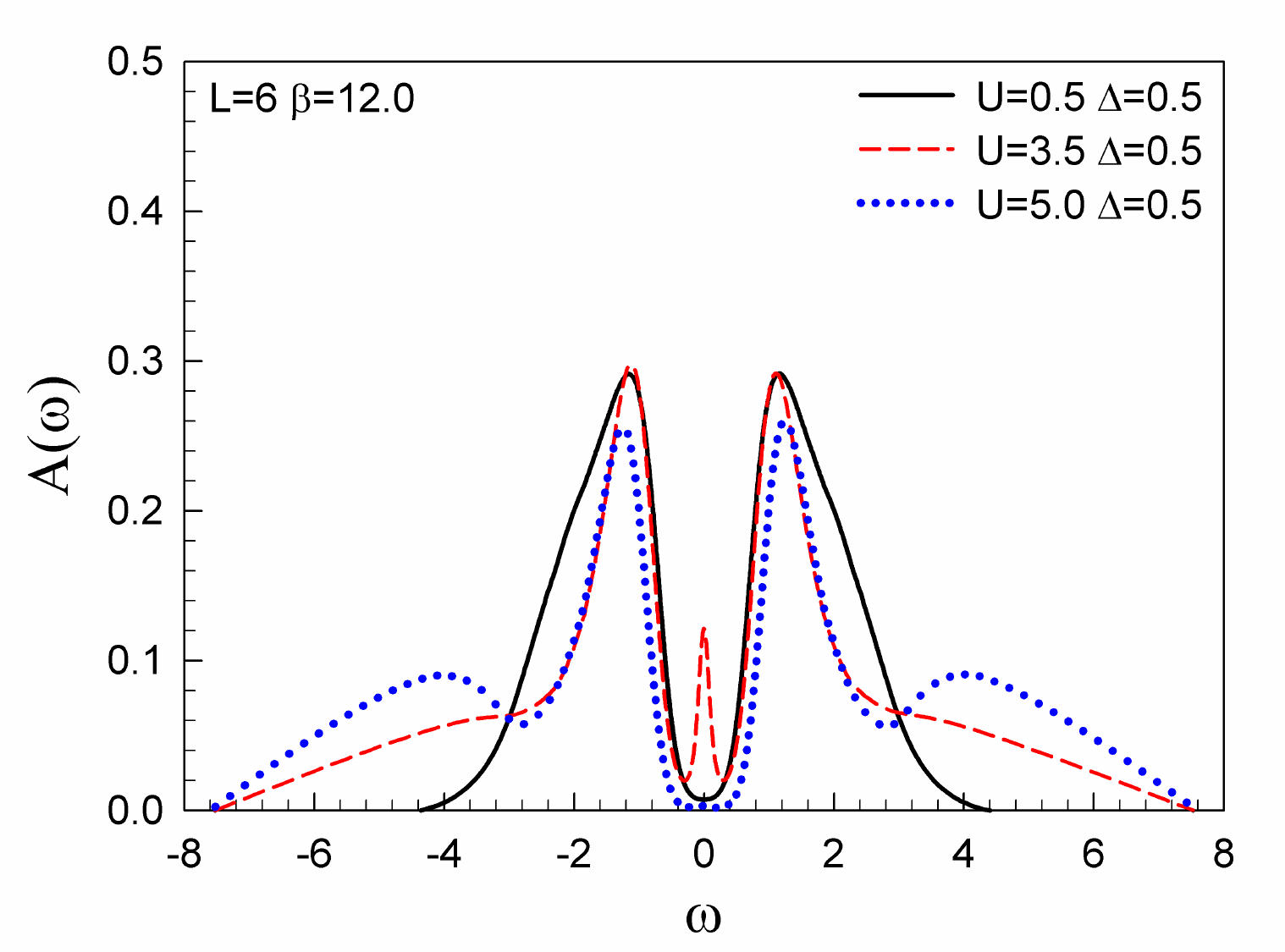}}
\caption{Spectral function $A(\omega)$ for different values of the on-site interaction strength $U$ with $\Delta$=0.5 for $L$=6 at $\beta$=12.0. The black solid line and the blue dotted line exhibit band and Mott insulators respectively, both of which show clear gaps at $\omega$=0. The red short-dashed line exhibits a metallic behavior as $A(\omega)$ is nonzero at $\omega$=0.}
\label{Fig:SpecFn}
\end{figure}

To further support our analysis of the intermediate phases shown in Figs.\ref{Fig:Phase} to \ref{Fig:U}, we calculated the spectral functions $A(\omega)$ for $\Delta$=0.5 and for three values of $U$ corresponding to the three phases. As shown in Fig.\ref{Fig:SpecFn}, for $U$=0.5 and $U$=5.0, corresponding to the band insulating and Mott insulating phases respectively, $A(\omega)$ shows a gap around $\omega$=0, thus indicating an insulator. In contrast, $A(\omega)$ for $U$=3.5 is nonvanishing and shows a quasiparticle peak at $\omega$=0, thus indicating a metallic phase. The results of spectral functions are consistent with the measurements of conductivity, as $U$ gets larger for $\Delta$=0.5, the spectral function shows an interaction-induced closing of a band gap and a subsequent opening of a Mott gap.

\begin{figure}[t]
\centerline {\includegraphics*[width=3.4in]{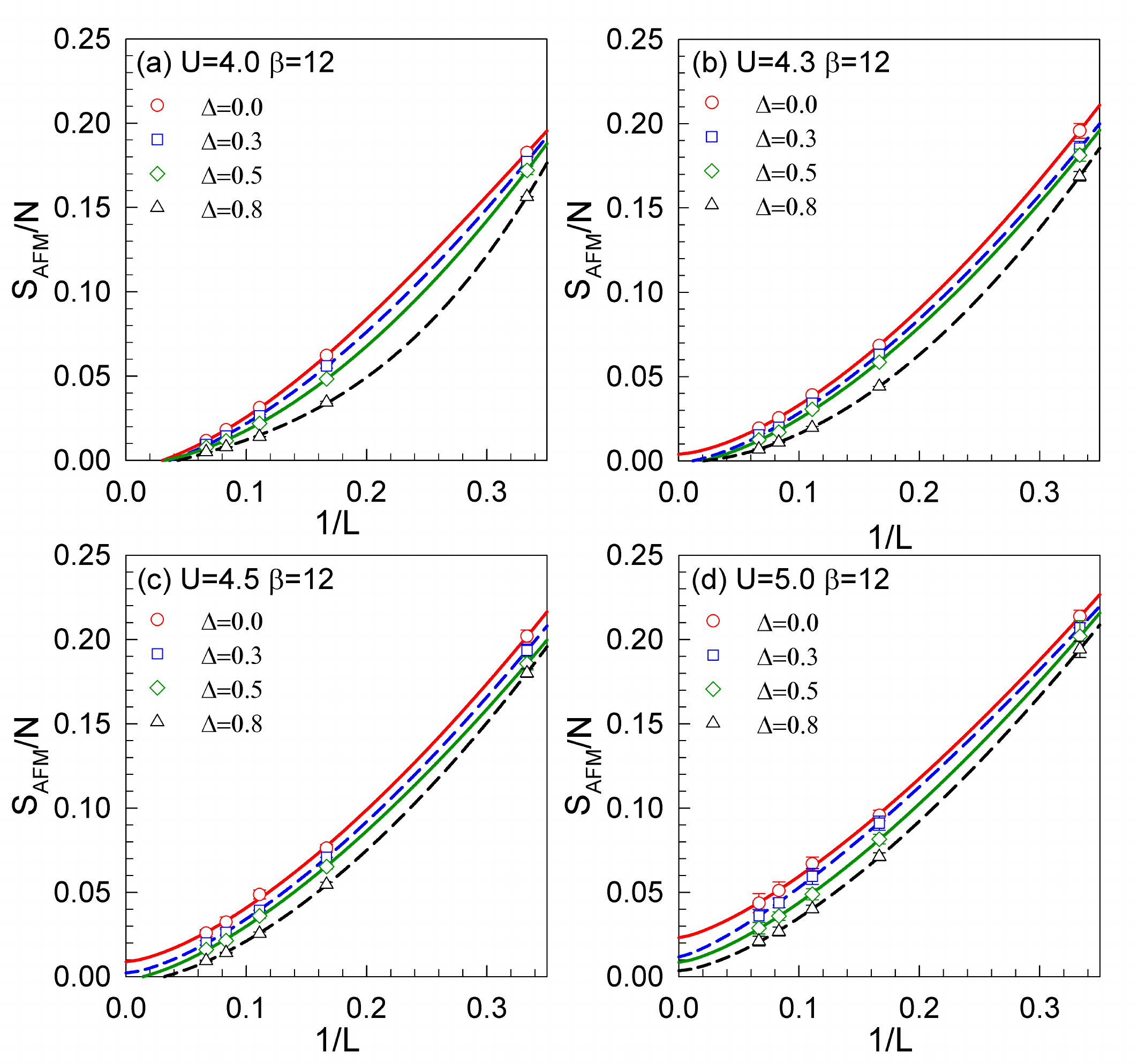}}
\caption{The AFM spin structure factor $S_{AFM}$ in the thermodynamic limit, which is plotted as a function of $1/L$ for different staggered potential values at fixed (a) $U$=4.0, (b) $U$=4.3, (c) $U$=4.5, and (d) $U$=5.0 at $\beta$=12.0. The scattered symbols are our AFM calculation results. The curves represent the results of cubic polynomial data fitting. A finite y-axis intercept in the $L$$\rightarrow$$\infty$ limit indicates that the long-range AFM order exists.}
\label{Fig:Spin}
\end{figure}

Fig.~\ref{Fig:Spin} provides the finite-size scaling results of the AFM structure factor on lattices of size $L$=3, 6, 9, 12, 15.
By extrapolating the data to the thermodynamic limit, we estimate that the critical point for $U$ is $U$=4.0$\sim$4.3 when the band gap $\Delta$=0, which is consistent with the previous studies of AFM order\cite{Sorella2012,*PhysRevX.6.011029}.
As we can see, $\Delta$ suppresses the AFM structure factor, while $U$ plays an opposite role. An increase in the on-site interaction to $U$=4.5 AFM order for $\Delta$=0.3, and a further increase to $U$=5.0 enables all calculated $\Delta$ values to exhibit AFM order.

\begin{figure}[t]
\centerline {\includegraphics[width=3.4in]{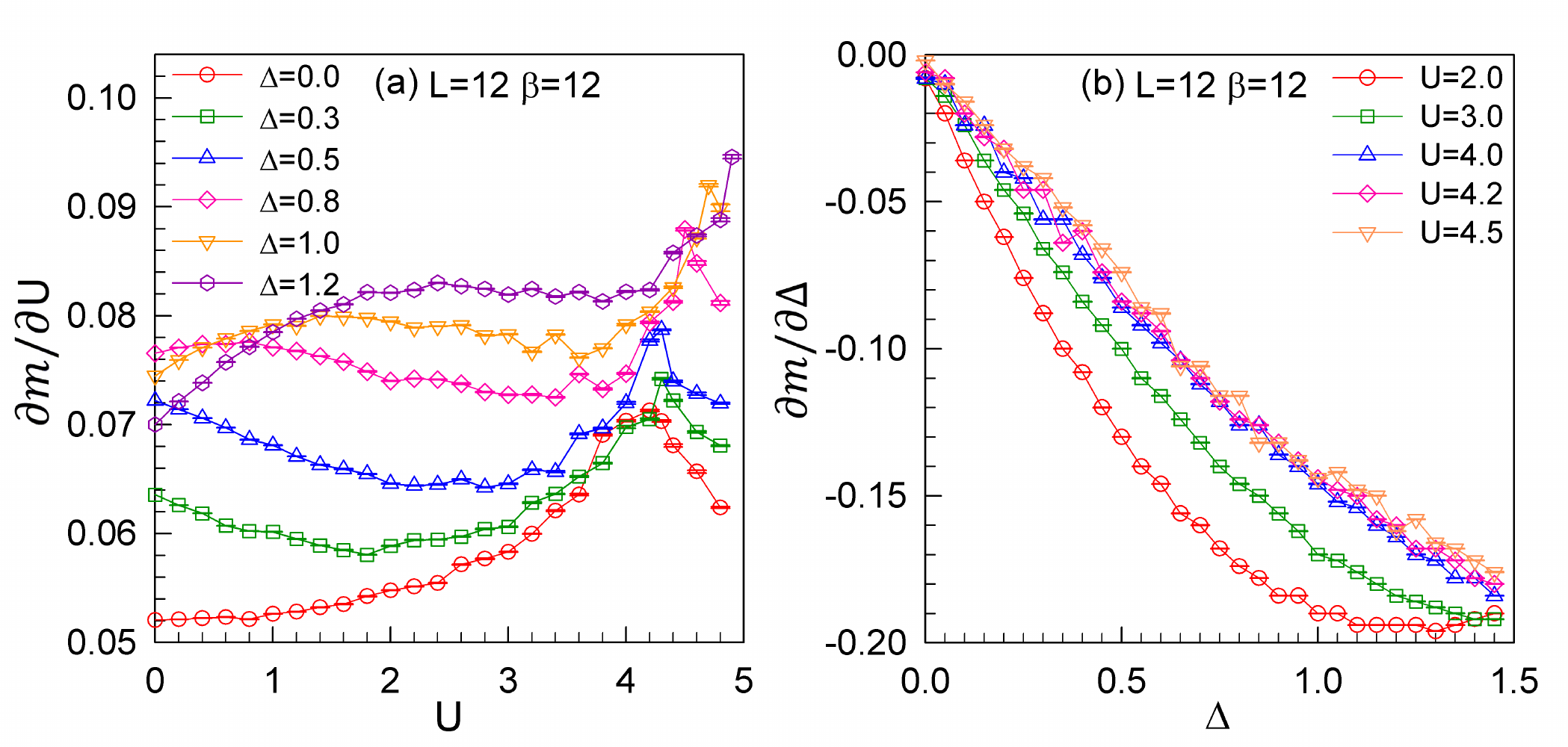}}
\caption{(a)First derivative of the local moment $m$ with respect to $U$ as a function of $U$ for fixed different values of $\Delta$, $\partial m/\partial U$, shows a local maximum value at the transition point from metal to Mott insulator state for small $\Delta$. (b)First derivative of $m$ with respect to $\Delta$, $\partial m/\partial \Delta$, as a function of $\Delta$ for fixed different $U$.}
\label{Fig:local moment}
\end{figure}

Local moment formation has been reported to happen discontinuously for the onset of Mott insulator behavior \cite{PhysRevB.87.125141, PhysRevB.80.155116, doi:10.1143/JPSJ.77.093703}.
Fig.\ref{Fig:local moment} (a) shows $\partial m/\partial U$ on $U$ for a range of fixed values of $\Delta$ for $L$=12 and $\beta$=12.
We find that the Mott metal-insulator transition is associated with a maximum value in $\partial{m}/\partial U$. The value of $U$ where the maximum in $\partial{m}/\partial U$ appears is very close to the $U_c$ characterized by the conductivity shown in Fig.\ref{Fig:Phase}.
 Although the error bars make the $U$ value of $\partial{m}/\partial U$ transition point somewhat uncertain, the discontinuity of $\partial{m}/\partial U$ is an indicator of the phase transitions.
The behavior of $\partial{m}/\partial \Delta$ versus $\Delta$ is given in Fig.\ref{Fig:local moment} (b). At small interaction ($U$=2.0, 3.0) the system undergoes a very robust metallic phase, then immediately changes into a band insulator phase as $\Delta$ gets larger, thus distinguishing the corresponding $\partial m/\partial \Delta$ lines from the large $U$ ones, as at large $U$ the phase transition of the system falls near the Mott metal-insulator boundary.

The local moment is related with the double occupancy $d$ of $m$=$1-2d$, and the evolution of $d$ may explain the phase transitions of the system to some extent.
At $T$=$U$=0 and $\Delta$$>$0, the system is an insulator with some double occupancy because of the gap. If $U$ is increased, the charges would rearrange to reduce the double occupancy. If increased enough, eventually there is an elimination of double occupancy, one spin 1/2 occupies each site, and the model has long-range AFM order.

We also compared our results with those of an experimental study\cite{PhysRevLett.115.115303}, in which their MI correlation image lies in our MI phase and CDW ordered phase lies in our band insulator (BI) phase. Their noise correlation image of the metal phase is a bit outside our metallic phase (which ends at $U_c$=3.9). However, our value for $U_c$, based on the conductivity, is consistent with the very precise simulations of Otsuka et al.\cite{Sorella2012,*PhysRevX.6.011029} which finds $U_c$=3.9 based on the magnetic structure factor.
Beyond this consistency, we obtain a phase diagram with numerically exact phase boundaries, making our results surpass the experimental work.

\begin{figure}[tbp]
\centerline {\includegraphics[width=3.4in]{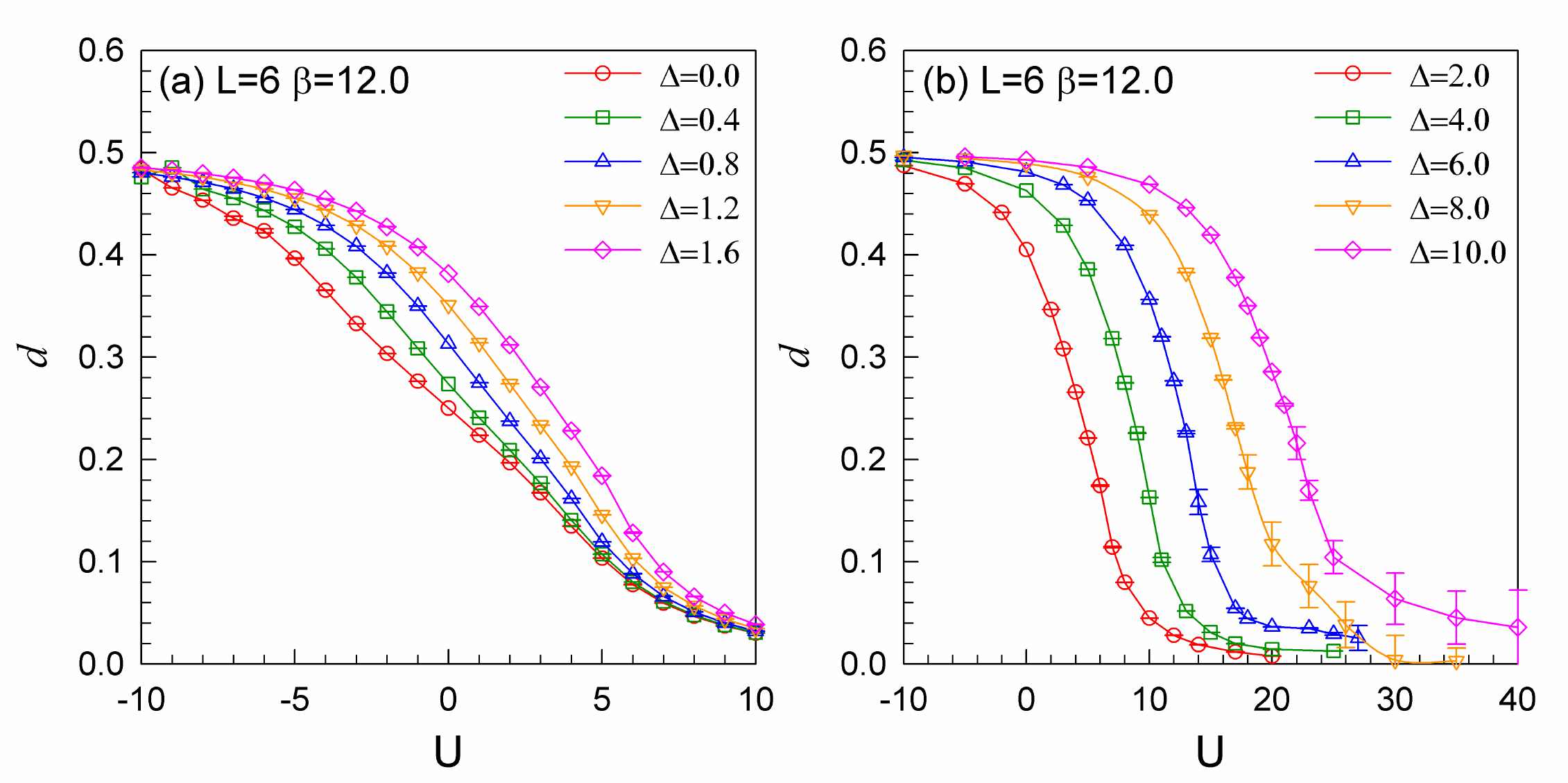}}
\caption{The double occupancy $d$ as a function of the on-site interaction $U$ for various values of $\Delta$ for $L$=6 at $\beta$=12.}
\label{Fig:doccu}
\end{figure}

To make a more direct comparison with the experiments\cite{PhysRevLett.115.115303}, the double occupancy $d$ as a function of the on-site interaction $U$ for various values of $\Delta$ is shown in Fig. \ref{Fig:doccu}.
For strong attractive interactions, a large fraction of doubly occupied sites is observed and it continuously decreases as $U$ increases.
In the weak repulsive interactions where $\Delta\gg U$, the double occupancy continues to be large, as is that expected for the band insulator.
For the strong interaction region of $U \gg\Delta$, the double occupancy tends to vanish for the largest $U$, consistent with a Mott insulating phase.
These trend and physical characteristics of double occupancy basically match the experimental results shown in Fig.2 of Ref. \cite{PhysRevLett.115.115303}.

\section{Conclusions}
\label{sec:conclusions}

%\noindent
%\underline{\it Conclusions}---
Between the band insulating phase and Mott insulating phase, we find a metallic phase in the ionic Hubbard model on a honeycomb lattice. On a square lattice\cite{PhysRevLett.97.046403,PhysRevLett.98.046403,PhysRevB.76.085112}, Coulomb interactions produce an AFM insulating phase at infinitesimal $U$, and the competition with the staggered potential induces a metallic phase\cite{PhysRevB.76.085112}.
By contrast, on the honeycomb lattice, a Mott insulator transition occurs at finite $U$ even without staggered potential, and a rather wide region of $U$ and $\Delta$ for the metallic phase is found.

In summary, we studied the ionic Hubbard model on a honeycomb lattice by a determinant quantum Monte Carlo method. We found that the intermediate phase between the two insulating phases was metallic. The staggered potentials drive the metallic phase to the band insulating phase, while the effect of the Coulomb repulsion was quite different. The Coulomb repulsion first drives the metallic phase into a nonmagnetic Mott insulating phase and then to the antiferromagnetic Mott insulating phase. As the Coulomb repulsion $U$ increases, the critical value of the staggered potential $\Delta _c$ increases, suggesting competition between the two energy scales. However, along the Mott metal-insulator transition line, the effect of staggered potential is weak as the electrons are localized in the Mott insulating phase.
Compared with previous theoretical proposals on some other models,
our extensive numerical studies succeed in achieving a complete phase diagram,
where the intermediate state is robust and occupies a large part of the phase diagram, and thus it should be more feasible to be detected experimentally.

%\acknowledgements
%\noindent
%\underline{\it Acknowledgment}---
\begin{acknowledgments}
We thank Richard T. Scalettar and James E Gubernatis for many helpful discussions.
We thank Hui Shao, Jinghua Sun and Tang Ho Kin for helping us to calculate spectral functions.
This work was supported by NSFC (No. 11774033, No. 11974049, No. 11974048 and No. 11504023) and Beijing Natural Science Foundation (No. 1192011).
We acknowledge the support of HSCC of Beijing Normal University, and some numerical simulations
in this work were performed on
Tianhe in Beijing Computational Science Research Center.
\end{acknowledgments}

\appendix

\section{Fintie size effect}
\label{appendix:A}

To make the phase diagram more accurate and convincing, order parameters computed on finite-size lattices must be extrapolated to the thermodynamic limit. The finite-size effect on the spin structure factor $S_{AFM}$ was carefully examined in the manuscript. Here we discuss the size effect of the DC conductivity.

\begin{figure}[htbp]
\centerline {\includegraphics[width=3.4in]{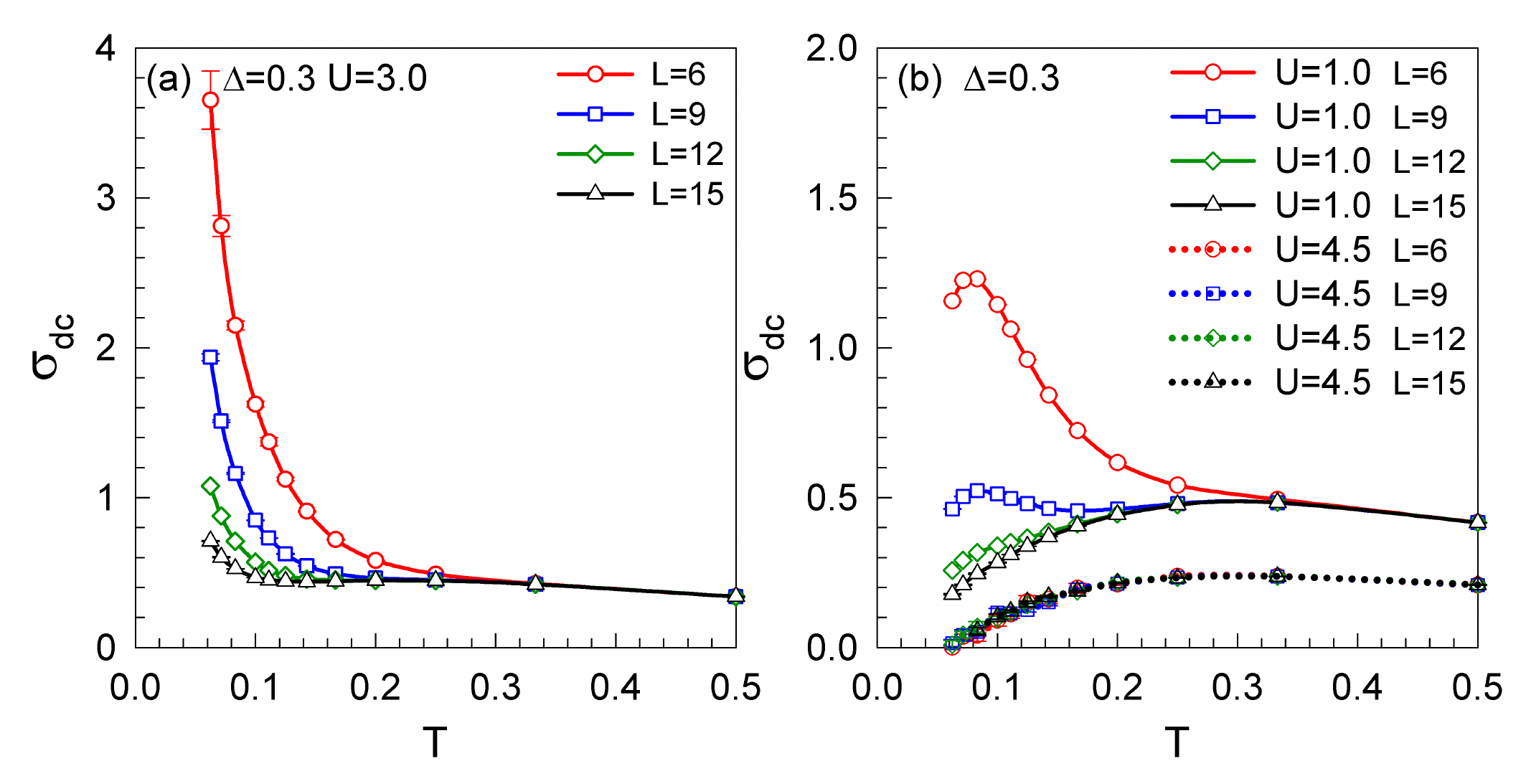}}
\caption{The conductivity $\sigma_{dc}$ is shown as a function of temperature at half-filling for various lattice sizes with  $\Delta=0.3$ of (a) metallic and (b) insulating states.}
\label{Fig:DcL}
\end{figure}

In Fig.~\ref{Fig:DcL}, we plot the conductivity $\sigma_{dc}$ as a function of the temperature for lattices up to $L=15$ at metallic states (a) and insulating states (b). Both Fig.~\ref{Fig:DcL} (a) and (b)($U$=1.0) indicate that the lattice size has a distinct influence on the conductivity for $U\leq3.0$.
This result is predictable because the finite-size effects have remarkable impact on weak coupling.
At $U=3.0$ and $\Delta=0.3$, there is an increase in $\sigma_{dc}$ with decreasing $T$ for the lattices that we have studied. Additionally, the metallic behavior weakens as the lattice size is increased. Although $\sigma_{dc}$ decreases with increasing lattice sizes, the signature of metallic behavior $d\sigma_{dc}/dT<0$ is unchanged. At $U=1.0$ and $\Delta=0.3$, the system shows an insulating behavior at low temperature, and results on larger lattice sizes reconfirm this behavior.
At larger interaction strength as $U=4.5$ for the insulating states, the conductivity is almost independent of the lattice size.

These findings are consistent with the
consensus that in a gapped system, one expects finite-size effects to be much smaller than in a metallic one.
Because our focus is to discern the insulating phase, the
data suggest that the $L = 12$ lattice is large enough to
be simulated for $\sigma_{dc}$, and we could ascertain
the insulating phase at low T.

\section{Zero temperature limit}
\label{appendix:B}

\begin{figure}[htbp]
\centerline {\includegraphics[width=3.4in]{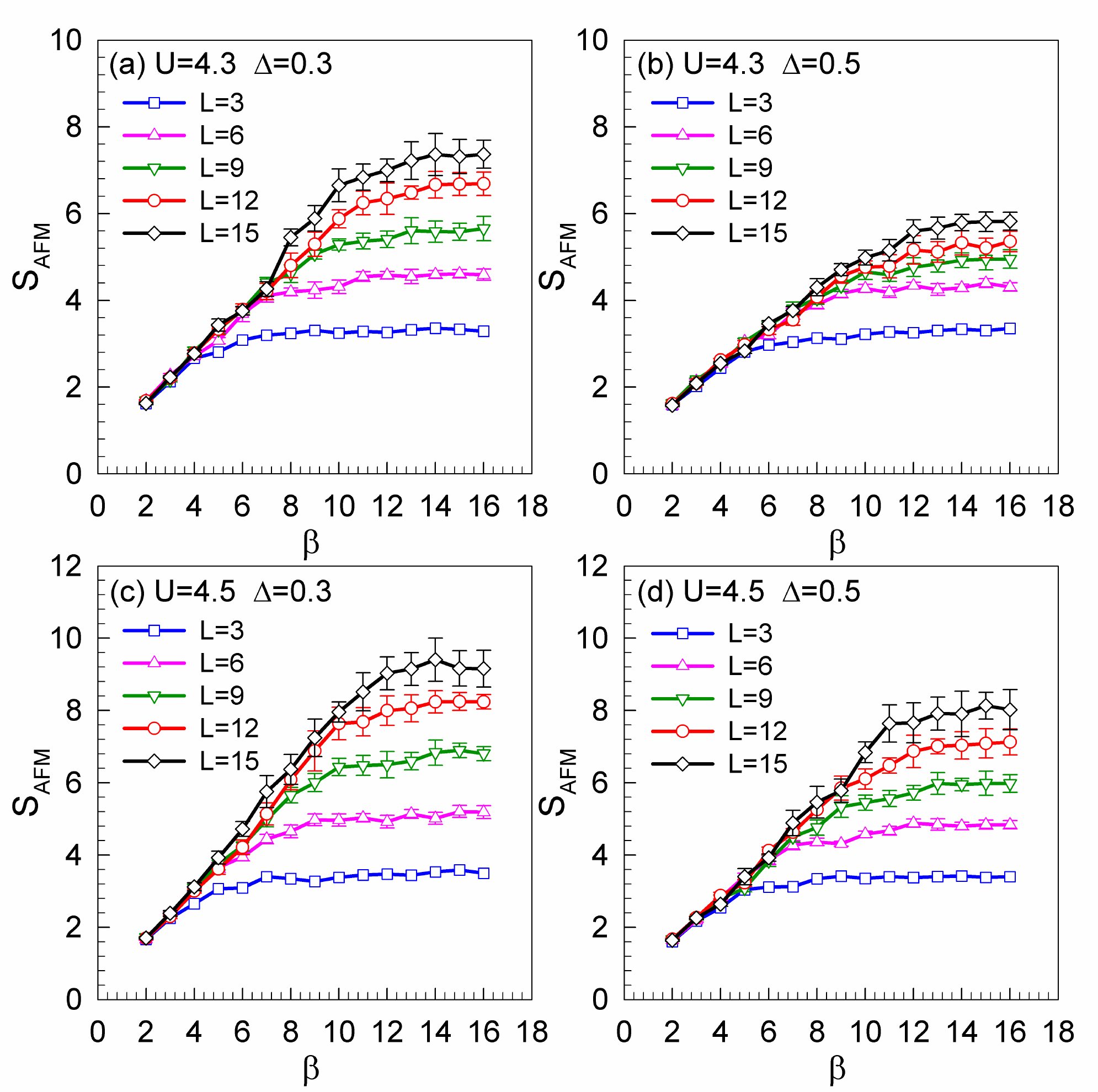}}
\caption{Temperature ($\beta$=1/T) dependence of the AFM spin structure factor $S_{AFM}$. At sufficiently low temperatures, $S_{AFM}$ saturates and becomes temperature independent within acceptable statistical errors.}
\label{Fig:Safm-beta}
\end{figure}

The numerical method we employ, the finite temperature determinant quantum  Monte Carlo (DQMC) method, can only calculate results at finite temperature. But we have ensured that the numerical results are converged at sufficiently low temperature, low enough to be regarded as the ground-state properties (zero temperature). Fig.\ref{Fig:Safm-beta} is the detailed example. In Fig.\ref{Fig:Safm-beta}, we plot the AFM spin structure factor $S_{AFM}$ as a function of the inverse temperature $ \beta = 1/T $ under circumstances of different lattice sizes and staggered potentials.

The figure shows that the AFM order increases as the temperature is lowered. When $T$ drops below a lattice-dependent temperature, $S_{AFM}$ saturates and gradually level off, barely no $\beta$ dependence within acceptable statistical errors. So we reasonably conclude that the physical observable has reached the $T=0$ ground state if its value is convergent below some $\beta_0 \simeq 10$. In this work, we evaluated the limit to be $\beta_0 \simeq 12$ for the spin structure factor. All the data presented in our paper were acquired at temperatures lower than $\beta_0$. Therefore, the results we obtained can be regarded as the $T=0$ ground-state properties under the corresponding conditions. Considering this, we can do the $1/L\rightarrow 0$ size scaling.

\section{The DC conductivity formula}
\label{appendix:C}

In our work, the low temperature behavior of DC conductivity $\sigma_{dc}$ is used to distinguish metallic or insulating phases. We implemented the  approach proposed in the work by  Trivedi et al.\cite{PhysRevB.54.R3756}, which is based on the following argument. From the fluctuation-dissipation theorem,
\begin{eqnarray}
\label{conductivity}
\Lambda_{xx}(\textbf{q},\tau)=\frac{1}{\pi}\int d\omega \frac{e^{-\omega \tau}}{1-e^{-\beta\omega}}\text{Im}\Lambda_{xx}(\textbf{q},\omega)
\end{eqnarray}
where $\Lambda_{xx}$ is the current-current correlation function. While $\text{Im}\Lambda_{xx}(\textbf{q},\omega)$ could be computed by a numerical analytic continuation of $\Lambda_{xx}(\textbf{q},\tau)$ data, we instead here assume that $\text{Im}\Lambda_{xx}\sim\omega\sigma_{dc}$ below some energy scale $\omega < \omega*$. Provided the temperature T is sufficiently lower than $\omega*$, the above equation simplifies to
\begin{eqnarray}
\label{simpconduc}
\Lambda_{xx}(\textbf{q}=0,\tau=\frac{\beta}{2})=\frac{\pi}{\beta^2}\sigma_{dc}
\end{eqnarray}
which is Eq. (2) in the manuscript.

It has been noted that this approach may not be valid for a Fermi liquid\cite{PhysRevB.54.R3756}. In this situation, the characteristic energy scale is set by $\omega* \sim N(0)T^2$. The requirement $T<\omega*$ will never be satisfied. However, in the system we studied, the energy scale is set by the temperature-independent staggered potential strength $\omega* \sim \Delta$ so that Eq.(\ref{simpconduc}) is valid at low temperatures.

It is worth mentioning that the authors of Refs.\cite{PhysRevB.54.R3756, PhysRevLett.120.116601} that used the DC
conductivity formula were simulating disordered systems, whereas it still applies to a system without disorder.
In Eq.(\ref{conductivity}), if we set $\tau$ to its largest value, $\tau=\beta/2$, where $\beta$ is large, then the $e^{-\omega\tau}=e^{-\omega\beta/2}$ factor dies off rapidly and only small $\omega$ values contribute to the integral. More specifically, we expect a low frequency behavior where $\text{Im}\Lambda\sim\omega\sigma_{dc}$ are important, and we can substitute this in and do the $\omega$ integration. This yields the approximate formula Eq.(\ref{simpconduc}).
Notice that for a Fermi liquid, $\Omega\sim1/\tau_{e-e}\sim N(0)T^2$ so it is impossible to satisfy $T\ll\Omega$. Hence we cannot use Eq.(\ref{simpconduc}). We are only safe if there is another scale (scattering mechanism), like disorder, which sets $\Omega$. For example, if we have disorder of strength $V$ and $\Omega \sim V$ we can reduce $T$ to the point where $T\ll\Omega$ and use Eq.(\ref{simpconduc}).

Now, even though we do not have disorder, we do have an energy scale $V$ associated with the staggered potential $V(-1)^{l}n_l$. If we Fourier transform this, it does scatter the fermions but only between $q$ and $q+\pi$, whereas a random potential mixes all $q$.

\bibliography{reference}
\end{document}